\documentclass[journal]{IEEEtran}
\usepackage{amsfonts}% Physical Review B
\usepackage{amsmath}
\usepackage{amssymb}
\usepackage{array}
\usepackage{mathrsfs}
\usepackage{leftidx}
\usepackage{graphicx}
\usepackage{epstopdf}
\usepackage{graphicx}% Include figure files
\usepackage{dcolumn}% Align table columns on decimal point
\usepackage{bm}% bold math
\usepackage[usenames]{color}
\usepackage{colortbl,booktabs}
\usepackage{subfigure}
\usepackage{cite}
\usepackage{multirow}
\usepackage{graphics}
\usepackage{stfloats}
\begin{document}
%
% paper title
% can use linebreaks \\ within to get better formatting as desired
% Do not put math or special symbols in the title.
\title{Networked Control Systems Secured by \\ Quantum Key Distribution}
\author{Hai-Jin Ding,
Re-Bing Wu,~\IEEEmembership{Member,~IEEE}, and Qian-Chuan Zhao
%\thanks{This work was supported by BNRist and National Natural Science Foundation of China (Grant Nos. **).}
\thanks{This work was supported by National Natural Science Foundation of China under Grant 61773232, 61374091 and 61425027, and the National Key Research and Development Program of China under Grant 2017YFA0304300. (Corresponding author: Re-Bing Wu)}
\thanks{The authors are all with the Department of Automation,
Tsinghua University, and also with the Beijing National Research Center for Information Science and Technology, Beijing 100084, China
(e-mail: rbwu@tsinghua.edu.cn; dhj17@tsinghua.edu.cn; zhaoqc@tsinghua.edu.cn ).}
%\thanks{Digital Object Identifier 10.1109/TCST.2012.2222883}
}

%% The paper headers
%\markboth{Journal of IEEE TRANSACTIONS ON CONTROL SYSTEMS TECHNOLOGY,~Vol.~11, No.~4, January~2014}%
%{Zhang\MakeLowercase{\textit{et al.}}: Minimum-Time Selective Control of Two Homonuclear Spins}
\maketitle

%%%%%%%%%%%%%%%%%%%%%%%%%%%%%%%%%%%%%%%%%%%%%%%%%%%%%%%%%%%%%%%%%%%%%%%%%%%%%%%%%%%
%%%%%%%%%%%%%%%%%%%%%%%%%%%%%%%%%%%%%%%%%%%%%%%%%%%%%%%%%%%%%%%%%%%%%%%%%%%%%%%%%%%
\IEEEtitleabstractindextext{}%
\begin{abstract}
%Cyber-security has become vital for modern networked control systems (NCS). In this paper, we introduce the emerging technology of quantum key distribution (QKD) to the protection of NCS privacy and security. QKD can safely distribute secret keys {\color{blue} of much higher privacy compared with traditional mesures} between communication parties, and thus enable the practical use of one-time pad encryption. %%%%{\color{blue}that is not realizable in classical communication}.
%We propose that the {\color{blue} overall} security of NCS can be essentially improved by one-time pad encryption using keys generated by QKD. Because the security is determined by the {\color{blue}quantum keys rather than the complexity of encryption algorithms,} the control performance is expected to be improved as well. In this regard, we present a composable definition of security based on the analysis of the key generation and the key management processes. Then, we analyze how the control performance of NCS can be influenced by the time delay induced by data encryption using various algorithms. With the example of remotely controlled servo, it is demonstrated that the introduction of QKD to NCS can simultaneously guarantee the security and performance by simplifying the encryption algorithms. Furthermore, we show that the combination of raw QKD with Kalman Filter can enhance available capacity of keys and privacy security without sacrificing too much on performance.

Cyber-security has become vital for modern networked control systems (NCS). In this paper, we propose that the emerging technology of quantum key distribution (QKD) can be applied to enhance the privacy and security of NCS up to an unbreakable level. QKD can continuously distribute random secret keys with much higher privacy between communication parties, and thus enable the one-time pad encryption that cannot be truly applied in classical networks. We show that the resulting overall security of NCS can be essentially improved, and present a composable definition of security based on the analysis of the key generation and management processes. Moreover, because the security is mainly determined by quantum keys rather than the complexity of encryption algorithms, the control performance can be improved as well by reducing the time delay using simpler algorithms. These advantages are demonstrated by the example of a remotely controlled servo system, showing that the introduction of QKD to NCS can simultaneously improve the security and performance by using the simplest encryption algorithm XOR. Furthermore, we propose a novel Kalman-filter embedded communication protocol that can more efficiently use the raw keys generated by QKD.
\end{abstract}

\section{Introduction}\label{Sec:Introduction}
\IEEEPARstart{C}YBER PHYSICAL SYSTEMS (CPS) are ubiquitous in modern industries~\cite{4519604}. As a typical application, many control actions are performed over CPS where the controller, the sensor and the plant are connected through high speed Ethernet or field bus. The resulting networked control systems can be flexibly designed based on time-driven or event-driven feedback, and the communication data, if necessary, can be stored and processed in the cloud~\cite{Kogiso2015,510638}. However, accompanied with the benefit is the high risk of cyber-security, because data protection is fragile or even absent in many traditional industrial control networks~\cite{Ding2017A}. Data transmission in such networks can be easily wiretapped or adversely changed by cyber attacks whose threat can be much severer than physical attacks~\cite{Kundur2011Towards}.

The cyber-security of NCS has drawn intense attention in the literature~\cite{Teixeira2010Networked}. According to the manner of attacks, cyber-attacks can be classified into denial-of-service (DoS) attack, replay attack, and deception attack. Various attack detection methods have been proposed ~\cite{Ding2017A}, such as Bayesian detection with binary hypothesis by testing with prior probabilities~\cite{Kailkhura2015Distributed} and weighted least square approaches by comparing the data of observer with a predetermined threshold~\cite{Deng2017Defending}. From a control theoretical point of view, the detection of cyber-attacks can be modeled and analyzed in terms of stability or  detectability problems~\cite{Shames2015A} and filtering methods can be applied to improve the performance of NCS\cite{WANG20162451}.

% \textbf{[MORE REFERENCES...]}

Comparing with control theoretic cyber-protection methods, strong data encryption is more straightforward and effective. Although data encryption is often restricted by limited (or sometimes unavailable) computation and communication resources, it is inevitable in critical infrastructures such as nuclear power plants where security is of the highest priority. Generally speaking, any encryption method involves an encryption algorithm for processing data and keys that are held by the users. The strength of data encryption, symmetric or asymmetric, is determined by the security of keys. However, because the keys cannot be frequently updated under most circumstances, almost all encryption methods used so far have to be strengthened by high algorithmic complexity (e.g., AES, DES, IDEA, RC2, RSA)~\cite{Mathur2015Solving}. Such algorithms either use long keys to form large-size cipher blocks, which increase the burden of communication, or use short keys to form small-size cipher blocks, which releases the communication burden but increases the complexity of computation~\cite{Mohd2015A,Deng2017A}.

One can certainly develop new encryption algorithms with higher complexity, but they are still under risk with more powerful computing technologies, not to mention the increased communication and computation burden. Especially, many currently used encryption algorithms (e.g., RSA) can be easily cracked by a quantum computer that is expected to be realizable in foreseeable future~\cite{Bernstein2017Post}. Under this circumstance, one may have to resort to one-time pad (OTP, i.e., each key is used only once) because of its provable absolute security that does not rely on the mathematical complexity of decryption algorithms \cite{MR0032133,Schneier1997Applied}. However, OTP is usually impossible over classical networks because the distribution of the unlimited amount of random keys between communication parties poses an equally difficult problem as the secure data transmission.
%~\cite{Mohd2015A} light block cipher

%\cite{Deng2017A} light key

%\cite{Lavanya2017LWDSA} light

Nevertheless, the obstacle that hinders the application of OTP can be overcome by the emerging technology of quantum key distribution (QKD)~\cite{Bennett1984}. In the most well-known BB84 protocol, random keys can be continuously generated and shared between communication parties using single photons (i.e., a socalled quantum channel). In terms of Shannon's~\cite{MR0032133} definition on perfect secrecy, it can be proved that BB84 is absolutely secure~\cite{476316,Shor2000Simple,Scarani2008Quantum,Benor2005The}. Moreover, a remarkable advantage of QKD is that malicious attacks on the quantum channel can be detected because of quantum no-cloning property~\cite{119991887739}.

%The security and performance of networked control system under cyber attacks is an important problem worth considering and it is meaningful to reform the performance with controlling method such as smoothing and filtering.
%Hongtao Sun et al.~\cite{Sun2017Resilient} analyzed the resilient control of networked control system under DoS attack with Lyapunov method and LMIs technique.
After two decades' intense research and development, QKD system has become commercially available. Demonstrated applications have appeared in financial and government fields~\cite{Kollmitzer2010Applied} in which security is highly demanded. To the authors' knowledge, no applications have been reported in networked control of cyber-physical systems where security is also critical. Despite of the potential high cost that can be gradually reduced in the future, QKD can bring at least three benefits when being applied to CPS:
\begin{enumerate}
\item The privacy security of data transmission can be essentially enhanced to an unbreakable level that is impossible with any classical communication system.
\item Cyber attacks can be more easily detected based on OTP that is enabled by QKD, because any adverse operation on the cipher texts will cause uncontrollable changes that can be discovered when being decrypted with synchronized encryption keys.
\item Guaranteed by strongly secure keys, it is in principle unnecessary to use complex encryption algorithms to secure the data transmission. Without sacrificing the security, simpler algorithms can be used to reduce the burden of computation and communication, and hence improve the control performance.
\end{enumerate}
%(4) raw quantum keys without the process of error correction can be directly used to encrypt NCS by modeling the error bits as noise and combining with filtering methods, a quite mature technology in practical network~\cite{Chen2017Event},\cite{Moayedi2009Adaptive}, thus the privacy security of system can be better ensured by the much larger capacity of raw keys and one-time pad.

This paper will demonstrate the above benefits based on our proposed integrated system structure for the QKD-based NCS. To evaluate the influence of QKD to security, network throughput and NCS control performance, we present measures for quantifying the improved trade-off, then the control performance can be greatly improved without sacrificing security. In addition to the system design and performance analysis, we also develop a novel encryption algorithm that pertains to the data transmission in NCS, which saves the amount of secret keys and further improves the security level.

%Related to the third benefit, the best tradeoff between control performance and security can be formulated as a bi-objective optimal control problem~\cite{Yau2009An}. In~\cite{6095624}, the optimal tradeoff was achieved using coevolutionary algorithm to solve a bi-objective optimization problem. In \cite{Haleem2007Opportunistic}, the optimal tradeoff between security and throughput in wireless network is quantified  under various adversary attacks.

The remainder of this paper is as follows. In Section \ref{Sec:Structure}, we design a practical NCS based on network communication and QKD. In Section \ref{Sec:Analysis}, we introduce the basic confidential security issues, study the factors that influence security and give a definition of security for networked control systems. In Section \ref{Sec:NCS-Delay}, we analyze the induced time delay caused by network, which will further influence the performance of NCS. In Section \ref{Sec:Tradeoff}, we study the performance of NCS using different encryption algorithms and the tradeoff between security and performance. In Section \ref{Sec:Raw}, we propose a new filtering based security strategy directly encrypting with the raw quantum keys that contain error bits,  then the quantum keys can be used more efficiently. Finally, in Section \ref{Sec:Conclusion}, conclusions are made.

\section{QKD Secured Networked Control Systems}\label{Sec:Structure}
In this section, we will introduce the control and communication structure of a networked control system equipped with a QKD device. The key generation methods and relevant encryption algorithms will be introduced as well.

\subsection{Structure of Networked Control Systems} \label{Sec:SubStructure}
\begin{figure}[h]
	\centering
	\includegraphics[width=1\columnwidth]{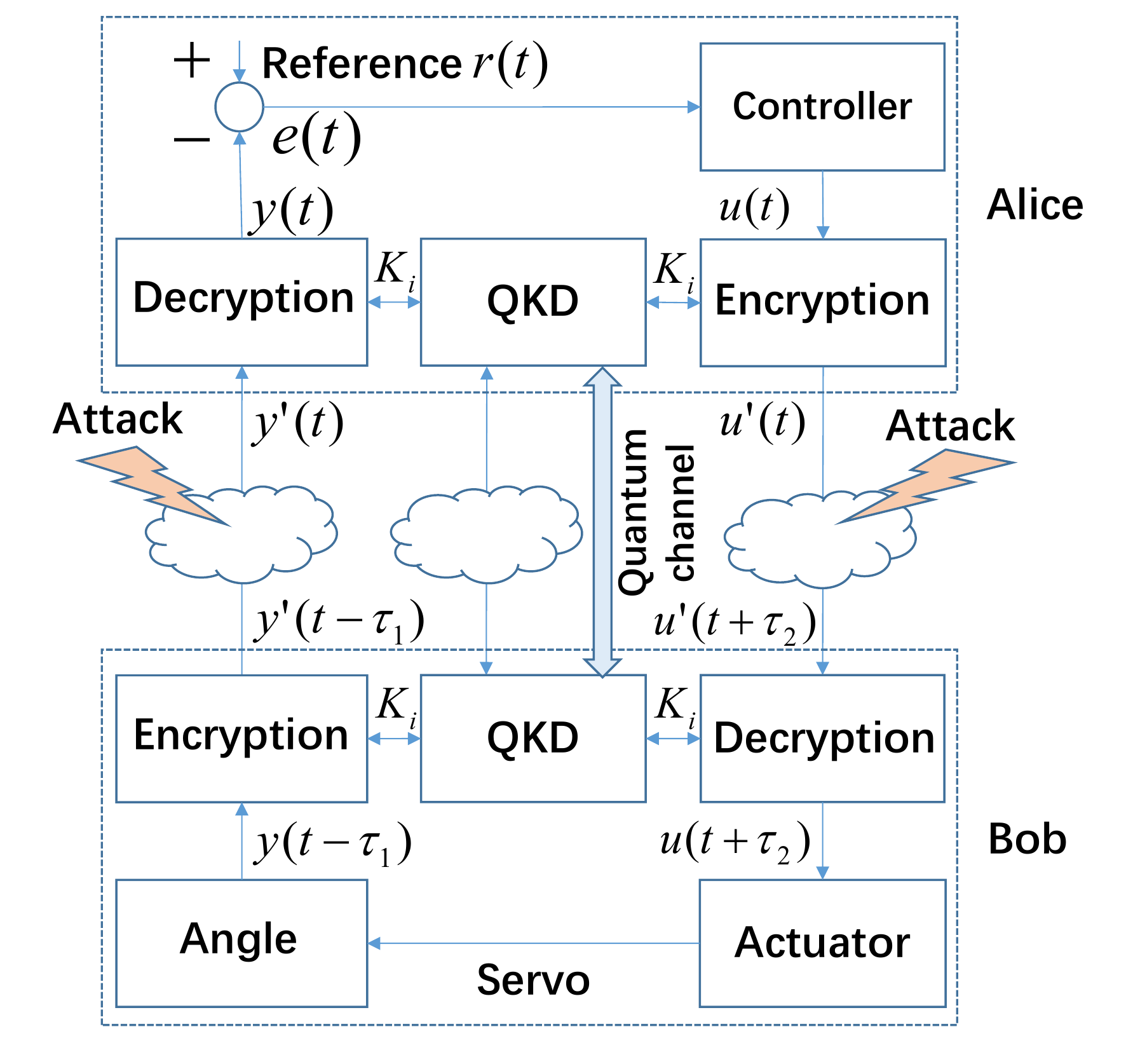}
	\caption{Schematic diagram of a QKD based networked control system.}
	\label{fig:topology}
\end{figure}

Consider a point-to-point NCS in which the plant is remotely controlled by a controller. For illustration, we use a mechanical servo as the example of plant. Any other dynamical control plants can be applied to the system introduced in the following. As illustrated in Fig.\ref{fig:topology}, the mechanical servo, Bob, is remotely controlled by Alice, a proportional-integral (PI) controller associated with another IP. Alice receives and decrypts the position data of the servo that is encrypted by Bob and sends back encrypted real-time control commands to Bob. The data transmission between Alice and Bob is based on communication protocol such as TCP/IP, UDP, etc.

To secure the data transmission, a QKD system is integrated into the network. It operates independently with the NCS, and supplies secret keys for the above data encryption and decryption of position data and control commands. The terminologies involved in this process are summarized in Table \ref{tab:notation}.

\begin{table}[htbp]
\centering
 \caption{Table of Terminologies}
 \begin{tabular}{  c  |  c}
  \toprule
$K_{i}$ & the $i$-th key used for encryption and decryption\\
$T_{K_{i}}$ & the encryption matrix of the key $K_{i}$\\
$T^{-1}_{K_{i}}$ & the decryption matrix of the key $K_{i}$\\
$u(t)$ &  the plain text of the control command sent from Alice\\
$u'(t)$ & the cipher text of the control command received by Bob\\
$y(t)$ &  the plain text of the position sent from Bob\\
$y'(t)$ & the cipher text of the position received by Alice\\
$\tau_{1}$ & the one-way delay from Bob to Alice\\
$\tau_{2}$ & the one-way delay from Alice to Bob\\
$r(t)$ & the reference signal\\
$e(t)$ & the error signal \\
  \bottomrule
 \end{tabular}
\label{tab:notation}
\end{table}

%$K_{i}$ ---- The i\textsuperscript{th} key used for encryption and decryption;

%$T_{K_{i}}$ ---- Encryption matrix of key $K_{i}$;

%$T^{-1}_{K_{i}}$ ---- Decryption matrix of key $K_{i}$;

%$m_{C}$ ---- The plain text of the control command sent from Alice;

%$C_{C}$ ---- The cipher text of the control command received by Bob;

%$m_{P}$ ---- The plain text of the position sent from Bob;

%$C_{P}$ ---- The cipher text of the position received by Alice;

In detail, Alice generates the control signal according to the error signal $e(t)$ as
\begin{equation}
e(t)=r(t)-y(t-\tau_{1}),
\end{equation}
where $r(t)$ is the reference signal and $y(t-\tau_{1})$ is the position signal of the servo sent from Bob after a time delay $\tau_{1}$. Assume that Alice adopts a PI controller, the generated control command is
\begin{equation}  \label{con:controlU}
u(t)=K_{P}e(t)+K_{I}\int_{0}^{t}e(t)dt,
\end{equation}
where $K_P$ and $K_I$ are the parameters of the controller.

Then, Alice encrypts the control command $u(t)$ with the key $K_{i}$, which produces the cipher text
\begin{equation}
u'(t)=T_{K_{i}}u(t),
\end{equation}
where $T_{K_{i}}$ represents the transformation posed by the encryption algorithm using the key $K_i$. The cipher text $u'(t)$ is then sent to Bob, which, after another time delay $\tau_{2}$, arrives and is decrypted using the key $K_i'$ as follows
\begin{equation}
u(t+\tau_{2})=T^{-1}_{K_{i}'}u'(t).
\end{equation}
Note that in symmetric encryption schemes, the keys $K_i$ and $K_i'$ are supposed to be identical. However, errors can occur during the distribution of keys between Alice and Bob (see Sec.~\ref{SubSec:BB84}), i.e.,
\begin{equation}
K_i'=K_i+\Delta_i,
\end{equation}
where $\Delta_i$ represents the difference between keys.

\begin{figure}[h]
	\centering
	\includegraphics[width=1\columnwidth]{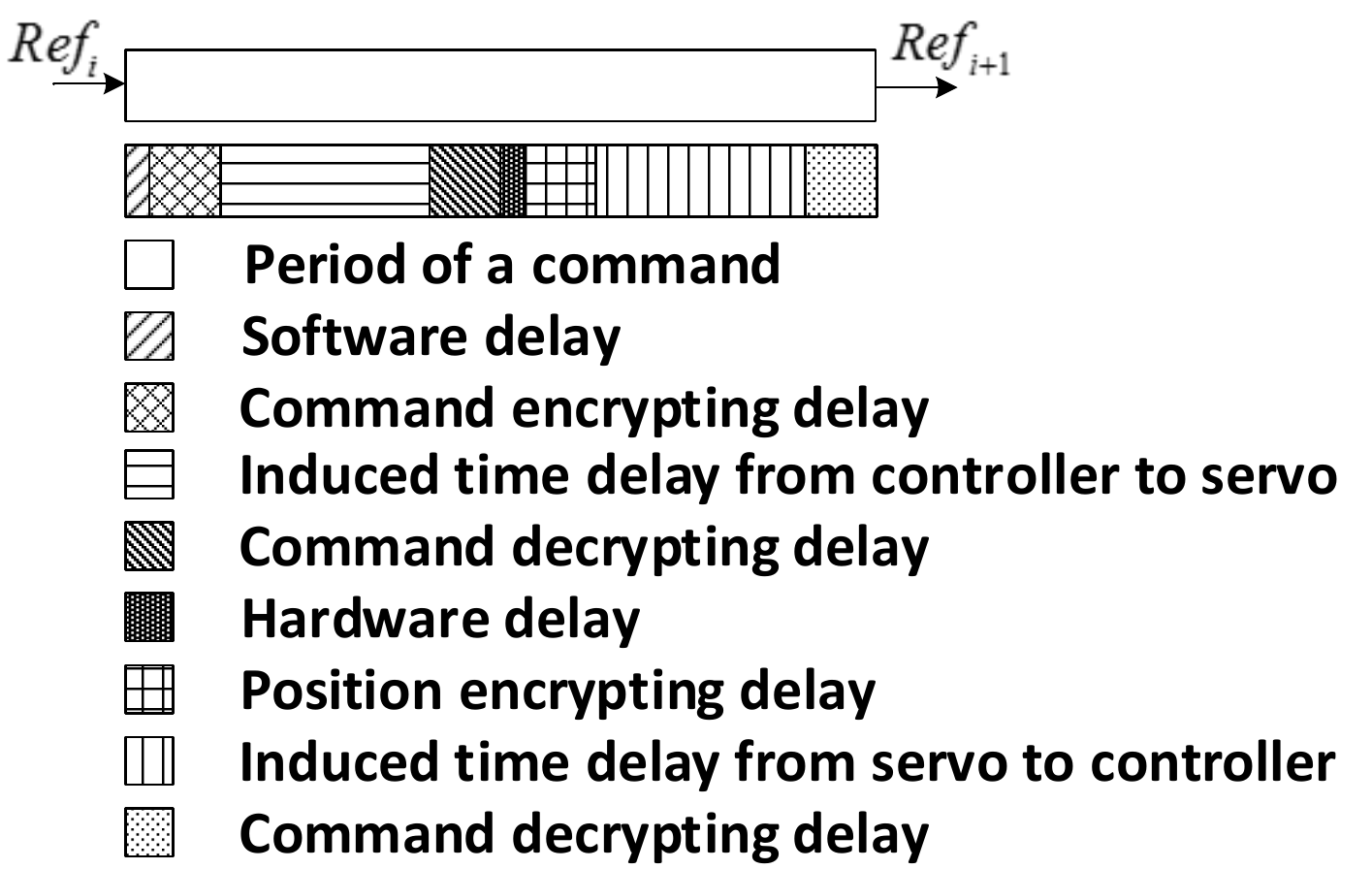}
	\caption{The time delays in a typical sampling control period of NCS.}
	\label{fig:delay}
\end{figure}

In practical networked control systems, the real-time performance is largely determined by the communication latencies of closed control loop. As is shown in Fig.~\ref{fig:delay}, the time delay includes software, hardware and network delays. So that the servo can only receive a command made according to its previous status which will deteriorate the performance.

%The closed-control period of the NCS can then be defined as the round-trip time delay, which is usually random. Consider the control process in the $i$-th control period, we summarize the following generalized state-space model for the encrypted dynamics of networked control systems:
% \begin{equation}  \label{con:model2}
%\left\{
%\begin{aligned}
%&x_{i}=Ax_{i-1}+Bu_{i-1}\\
%&y_{i}=Cx_{i},\\
%&u_{i}=F\cdot T^{-1}_{K_i+\Delta_i}T_{K_i}\cdot y_{i},\\
%\end{aligned}
%\right.
%\end{equation}
%where $x_{i}$ is the state vector, $u_i$ is the control command and and $y_i$ is the output. The first two equations   represents the standard linear dynamics and measurement of the plant. The last equation describes the process that the measurement data $y_i$ is encrypted (using key $K_i$) by Alice and decrypted by Bob (using key $K_i'$ with possible error $\Delta_i$). The received data is then processed by the feedback matrix $F$ to produce the next control command. The random key error $\Delta_i$ introduces a noise that can be attenuated by Kalman filter, which will be discussed later in Sec.~\ref{SubSec:Raw}.

The closed-loop control period of the NCS is defined as the round-trip time delay, which is usually random. Consider the control process in the $i$th control period, we summarize the following generalized state-space model for the encrypted dynamics of the networked control system:
% \begin{equation}  \label{con:controlmodel}
%\left\{
%\begin{aligned}
%&x_{i}=Ax_{i-1}+B\cdot T^{-1}_{k_{i-1}+\Delta_{i-1}}T_{k_{i-1}}u_{i-1}\\
%&y_{i}=Cx_{i},\\
%&u_{i}=F\cdot T^{-1}_{k_i+\Delta_i}T_{k_i}\cdot y_{i},
%\end{aligned}
%\right.
%\end{equation}

 \begin{equation}  \label{con:controlmodel}
\left\{
\begin{aligned}
&x_{i}=Ax_{i-1}+Bu_{i-1},\\
&\bar{y}_{i}=Cx_{i},\\
&u_{i}=T^{-1}_{k_{i}+\Delta_{i}}T_{k_{i}}\bar{u}_{i},\\
&y_{i}=T^{-1}_{k'_i+\Delta'_i}T_{k'_i}\bar{y}_{i},\\
&\bar{u}_{i}=F\cdot(r_{i}-y_{i}),\\
\end{aligned}
\right.
\end{equation}
% \begin{equation}  \label{con:controlmodel}
%\left\{
%\begin{aligned}
%&x_{i}=Ax_{i-1}+B\cdot T^{-1}_{k_{i}+\Delta_{i}}T_{k_{i}}u_{i-1}\\
%&y_{i}=Cx_{i},\\
%&u_{i}=F\cdot(r_{i}-T^{-1}_{k'_i+\Delta'_i}T_{k'_i} y_{i}),
%\end{aligned}
%\right.
%\end{equation}
where $x_{i}$ is the state vector, $\bar{u}_i$ and $\bar{y}_i$ are the original control and output of the plant, $u_i$ and $y_i$ are transmitted control and output after encryption and decryption, and $r_i$ is the reference signal. The first three equations represents the standard linear dynamics and measurement of the plant, in which the control command $u_{i-1}$ is received by the plant after encryption with key $k_{i}$ and decryption with key $k_{i}+\Delta_{i}$, where $\Delta_{i}$ represents potential errors in the key. The fourth equation describes the process that the measurement data $y_i$ is encrypted (using key $k'_i$) by Bob and decrypted by Alice (using key $k'_i+\Delta'_i$ with possible error $\Delta'_i$). The received data is then processed by the feedback matrix $F$ to produce the next control command.

%\begin{figure}[h]
%	\centering
%	\includegraphics[width=0.8\columnwidth]{delay2.pdf}
%	\caption{The relationship between sampling control period of NCS and delay.}
%	\label{fig:compare}
%\end{figure}

%While on the other hand, time sequence control model will enlarge the overshoot of performance because the time delay is much longer than the sampling period of hardware. We can solve the problem with PWM modulation by multiplying the command with a square wave function of a relatively small duty cycle to avoid the servo going unstable.

\subsection{Generation of symmetric secret keys with QKD}\label{SubSec:BB84}
The function of QKD is to safely and continuously generate symmetric keys between Alice and Bob. In this paper, we will use the BB84 protocol, which is the earliest and also the most broadly used protocol in practical systems.

The major difference of QKD from its classical counterpart is that the secrecy of generated keys is physically guaranteed by quantum mechanics. In detail, the protocol can be divided into the following parts.
\subsubsection{Key Sifting}
Alice prepares and sends a bit encoded by single photons that are randomly polarized along two sets of bases, each of which having two orthogonal directions. Bob receives a bit after measuring the received photon along one of the two bases he randomly chooses. According to quantum mechanics, the measurement outcome is deterministic if and only if it is performed along the right basis, otherwise the outcome is random. Therefore, Alice and Bob can tell each other through classical communication which basis they choose instead of the bits themselves, and keep those bits that are sent and measured under the same basis. These bits are thus symmetrically distributed between Alice and Bob, forming the set of raw keys for encryption and decryption.

\subsubsection{Error correction}
Because the raw keys may contain errors due to the imperfection of single photon detector or losses during the transmission, a Cascade protocol is used to locate the wrong keys by comparing the check bit and halving the key strings. This process is continued until the exact positions of wrong bits are found, which will be corrected or discarded.

\subsubsection{Privacy amplification}
To avoid the information leakage in the classical communication for basis exchange and error correction, a privacy amplification protocol is applied using a Hash function, which guarantees the absolute security of the final key in an asymptotic manner. As a price, the capacity of keys will be further reduced during this process.

%(1)QKD is much safer than traditional key generating methods and the information that eavesdropper can get after privacy amplification is exponentially small.

%(2)The keys are distributed directly in two separated ports instead of transported through the ethernet, that will help keep the keys away from being cracked by eavesdropper.

%(3)The physical random number generator used in QKD is the most random type can be generated in practical system, and that can help make sure the true randomness of key series.

%(4)QKD is a real-time generating method and we can adjust the key policy according to the capacity of key pool.

\subsection{Encryption Algorithms}

The security of a cryptographic system is essentially determined by that of secret keys. Ideally, the data transmission is absolutely secure if every key is used only once (i.e., one-time pad). However, because it is usually impossible to distribute unlimited number of keys over classical communication networks, one has to enhance the security by strong encryption algorithms.

We will briefly introduce some widely used symmetric encryption algorithms. These algorithms all belong to the category of block cipher, which divides the plain text into identical-size blocks~\cite{Tang2010Research}. These blocks are processed to generate the cipher text via a series of basic operations (e.g., ByteSub, ShiftRow, MixColumn and RoundKeyAddition, see details in~\cite{10.1007/978-3-540-39887-5_19}) using a secret key.

%In detail, the most frequently used algorithms are summarized as follows.
%Usually the plain text string can be transferred to a matrix by an initial permutation. ByteSub is accompanied by S-box, a reversible matrix that can substitute each byte of plain text to a new byte and its security is connected with the dimension of S-box. MixColumn can shift the bytes of each single row for different bytes and can be finished by multiplying a matrix. Based on this, we analyze the security of different encryption algorithms.

\subsubsection{XOR} the simplest method for encryption, which is performed by bit-by-bit XOR operation on the plain text and keys, namely the RoundKeyAddition.

\subsubsection{DES (Data Encryption Standard)} the algorithm based on rounds of Feistel operations. The standard DES consists of 16 Feistel units, each of which halves the $i$-th round text into $L_{i}$ and $R_{i}$, and updates them as:
\begin{equation}   \label{con:DES}
L_{i+1}=R_{i}, \quad
R_{i+1}=L_{i}\oplus F(R_{i},K_{i+1}),
\end{equation}
where $K_i$ is the sub-key used in this round and $F(R_{i-1},K_{i})$ is the result of last Feistel round.
The sub-keys used in each round is generated by the ByteSub operation from the original key, and a simple MixColumn is used during the exchange of left and right halves according to (\ref{con:DES}). DES is actually an extension of XOR, but the security is much higher owing to the complicated bit position operation.

\subsubsection{AES (Advanced Encryption Standard)} the algorithm based on rounds of operations that uses the Substitution-Permutation Network (SPN) to generate round keys. The text is processed by cycling bit operations in each round. AES is much more complex than DES because of the increased key length as well as more complex round function (See Tab.~\ref{tab:AlgorithmComparision}). Presently, AES(128) with 128-bit key and 10 SPN rounds, AES(192) with 192-bit key and 12 SPN rounds, and AES(256) with 256-bit key and 14 SPN rounds, are most popular.

\subsection{Advantages of QKD-Based NCS}
Before expanding our studies, we make some remarks on how QKD may improve the security against cyber attacks by enabling the OTP.

%{\color{blue} Let $\mathcal{E}$ be the set of admissiable plain texts (e.g., the range of control commands or sampled measurement data).}

%{\color{blue} To make sure the detection of attacks, the possible plain text is enlarged to be a combination of data set of communication (i.e., $u_{i}$, $y_{i}$ in (\ref{con:model2})) and another set of symbols $\mathcal{E}$ that never happens in an unattacked system.}

\subsubsection{DoS attack} in such attack, Eve blocks the transmission of cipher texts by maliciously overloading the communication of network. Although none of plain texts are cracked, the communication can be interrupted between the actuator and the controller and thus deteriorate the performance. Note that the DoS or the dropout can be detected with TCP/IP protocol by handshaking, but they may not be efficiently detectable with other protocols such as UDP.

When one-time pad and QKD are applied, it is easy to detect the occurence of DoS no matter which network communication protocol is used. This is because every data block Bob receives must be decrypted with the same key for encryption. Let $\mathcal{E}$ be the set of
admissible plain texts (e.g., the range of control commands or sampled measurement data). If some blocks are missing during the transmission, and Bob uses the key that is supposed to be used on the missing block to decrypt some other blocks he receives later, he cannot get a meaningful message because the key does not match, namely $T^{-1}_{K_i}(T_{K_{i+j}} u_{i+j})\notin\mathcal{E}$, where $K_{i}$ and $K_{i+j}$ are the desynchronized keys used for decryption and encryption due to DoS attack. In this way, DoS can be detected.
%Once DoS is detected, the control performance can be improved by event based feedback.

\subsubsection{Replay attack} in such attack, Eve replaces the transmitted messages from Alice to Bob with some previously recorded ones, in order to deceive Bob by the old messages that are still meaningful to him. When the messages are encrypted, the deception can be successful if the key is not changed, which is often the case in classical networks. However, it will not work when every key is used only once (i.e., one-time pad) because the replayed cipher text will not be decrypted by the right key that is only shared between Alice and Bob. Thus, when Eve performs the replay attack, Bob will not obtain a meaningful plain text as $T^{-1}_{K_{i+j} }(T_{K_{i}} u_{i})\notin\mathcal{E}$, where $K_{i}$ and $K_{i+j}$ are the desynchronized keys used for encryption and decryption due to the replay attack, and this can be used to detect the occurrence of the attack.
%When encrypted by a different key, which usually produces meaningless messages. Take ASCII table as an example, if the replayed cipher text $m_{i}$ is encrypted with XOR algorithm by $K_{i}$, copied at time $i$ and replayed at time $j$, namely:
%\begin{equation}
%m_{i}\oplus K_{i}=c_{i}.
%\end{equation}
%When the cipher text $c_{i}$ is replayed at time $j$, then the new plain text will be decrypted as follows:
%\begin{equation}
%c_{i}\oplus K_{j}=m_{j}^{*}.
%\end{equation}
%Ensured by random key generation, $K_{j}$ is totally different from $K_{i}$. Because the cipher text space $\mathcal{C}$ and plain text space $\mathcal{M}$ are large, the decrypted plain text $m_{j}^{*}$ is a random combination of elements from space $\mathcal{M}$ rather than a meaningful text, which is usually a messy code.

\subsubsection{Deception attack} in such attack, Eve injects malicious data in the communication to deteriorate the performance of NCS, e.g., by changing several bits of the cipher text. Similar to the case of replay attack, the injected data will not be able to convey a meaningful and spurious message if the cipher text is changed before decrypted by Bob as  $T^{-1}_{K_{i}}(T_{K_{i}} u_{i}+\Delta)\notin\mathcal{E}$, where $\Delta$ represents Eve's modification on the cipher text. Therefore, the deception attack can also be easily detected from the resulting messy code when OTP is enabled by QKD. One can also determine whether the attack is a replay attack or a deception attack by checking the repetition of the cipher texts.
% of the plain text is changed while the key $K_{i}$ is not changed. The difference between $m_{j}$ and $m_{j}^{*}$ is decided by the how many bits of $c_{i}$ is changed and the complexity of encryption algorithms.
%If the cipher text is modified as $c_{i}^{*}$, it will resulting in the following decryption:
%\begin{equation}
%m_{j}^{*}=c_{i}^{*}\oplus K_{i}.
%\end{equation}

\section{Security Measure of QKD-based networked control systems}\label{Sec:Analysis}

For the purpose of control systems analysis and design, a quantitative evaluation of the overall cyber-security is necessary. Generally speaking, the security of networked control systems is determined by the security of keys (i.e., how the keys are generated and distributed), the security of key management (i.e., how frequently every key is reused) and the security of the encryption algorithm (i.e., how hard the algorithm can be cracked).

Regarding these factors, we firstly define the security $S_A$ of an encryption algorithm. Suppose that the length of the key is $N$ and $R$ is number of rounds performed in the algorithm. We define the security strength of the encryption algorithm as follows (see the derivation in Appendix A):
\begin{equation}   \label{con:SA}
S_{A}=\eta R\log_{2}N,
\end{equation}
where $\eta$ represents the complexity of each round operation (e.g., the Feistel of DES or the SPN of AES). In Table \ref{tab:algorithm}, the security strength of several encryption algorithms are calculated according to the measure (\ref{con:SA}). Because AES(256) is recognized as the currently strongest symmetric encryption algorithm, we normalize the measure by that of AES(256).
\begin{table}[htbp]
\centering
 \caption{Security measure of various encryption algorithms}
 \begin{tabular}{  c  c  c  c  c }
  \toprule
  Algorithm & $\eta$ & $R$ &$\log_{2}N$ & $S_{A}$\\
  \midrule
 XOR & 0.0625 & 1 & 3 & 0.0017\\
 n-Feistel & 0.2083 & $n$ & 6 & 0.0112n\\
 DES & 0.2083 & 16 & 6 & 0.1785\\
 AES(128) &1 & 10 & 7 & 0.6250\\
 AES(192) &1 & 12 & 7.58 & 0.8121\\
 AES(256) &1 & 14 & 8 & 1\\
  \bottomrule
 \end{tabular}
\label{tab:algorithm}
\end{table}

Based on $S_A$, the overall security of the QKD-based NCS is quantified as follows:
\begin{equation}  \label{con:securitydefinition}
S=1-\varepsilon r (1-S_{A}),
\end{equation}
where $r$ is the average times ($r=1$ for OTP and otherwise $r>1$) for the keys to be reused before being discarded, and $\varepsilon$ represents the insecurity of the keys generated by QKD after privacy amplification. The insecuriy parameter $\varepsilon$ can be arbitrarily small under properly designed QKD protocol.

Equation (\ref{con:securitydefinition}) shows that the security can be arbitrarily strong when the keys are sufficiently safe, under which circumstance the strength of encryption algorithms is not as important as before.

\section{Round-trip Delay of QKD-Based networked control}\label{Sec:NCS-Delay}
%The networked control system mainly different from traditional control system in performance and security. The performance of NCS is influenced a lot by time delay. And it can be divided into four parts: software delay, encryption-decryption delay, transmission delay in Ethernet and hardware delay. Among the four parts of delay, encryption-decryption delay and transmission delay are the two major parts. While the transmission delay is decided by the bandwidth and velocity of Ethernet. While the encryption-decryption delay can be reduced by simplifying encryption algorithm.

The real-time performance of practical networked control systems can be seriously deteriorated or even destabilized by the time delay of the closed loop control over the communication network. In this section, we analyze and quantify this effect, especially how it is affected by data encryption and decryption processes.

%\begin{figure}[h]
%	\centering
%	\includegraphics[width=0.8\columnwidth]{delay2.pdf}
%	\caption{The relationship between sampling control period of NCS and delay.}
%	\label{fig:compare}
%\end{figure}

%While on the other hand, time sequence control model will enlarge the overshoot of performance because the time delay is much longer than the sampling period of hardware. We can solve the problem with PWM modulation by multiplying the command with a square wave function of a relatively small duty cycle to avoid the servo going unstable.

%The elements of the round-trip time delay is shown in Fig.\ref{fig:delay}.
The end-to-end (one-way) delay can be decomposed into four types of delays: \emph{software delay} for the generation of the control command, \emph{encryption-decryption delay}, \emph{network delay} for data transmission and \emph{hardware delay} caused by the plant and other devices in the system. Among these elements, the encryption-decryption delay is dependent with the complexity of the adopted algorithm, the transmission delay varies according to the communication load as well as available bandwidth, and the other delays are relatively short and fixed during the control process.

We are mainly concerned with the time delay caused by data encryption and decryption, which is controllable by the choice of algorithms. Highly complicated encryption algorithms may increase the time for computation, but such difference can be minor under powerful computational hardware. The more important factor is the size of the cipher text blocks that directly influences the communication load of network. Therefore, the time delay caused by encryption-decryption is essentially determined by the length of the key that is proportional to the size of the cipher text blocks. For example, AES(256), one of the most secure algorithms, uses 256-bit keys.

In the QKD-based NCS, keys are important resources. To guarantee that there are always sufficient secure keys for one-time pad encryption, the consumption rate of keys by NCS must be lower than the generation rate of keys by QKD. Complex encryption algorithms usually generate long cipher texts using long keys. Consequently, more communication is needed for generating such keys in BB84, which induces larger time delays. This factor is important in the tradeoff between security and performance.

%of higher complexity will add the length of cipher text and the amount of bits needed to be transferred to finish error correction in BB84, so the induced time delay caused by communication load will be added. Based on this we study the relationship between security and performance.

%We define $\tau$ as the loop delay caused by classical communication between Alice and Bob. It is a combination of the delay caused by transferring cipher text and the classical communication when generating keys.

%The total bits transferred between Alice and Bob can be calculated as $l=N+l_{EC}$, where
Let $N$ be the length of cipher text and $l_{EC}$ be the number of bits that must be exchanged for error correction. In the Cascade protocol of BB84, $l_{EC}$ varies with $N$ and quantum bit error rate (QBER) $e$. Practically, QBER is evaluated by checking some bits of raw keys at the beginning, and it asks the QBER must be less than $11\%$~\cite{Shor2000Simple}. Then the keys are rearranged and divided into identical-size blocks so that there is only one error bit in each block. Let $l_{0}$ be the length of block, which is required to be no less than $[\frac{1}{e}]$. %Take one block as an example and there is only one error bit in the block.
%Alice and Bob independently calculate the check bit of keys and compare the results with each other, which will be different because the existence of error bit. Then both Alice and Bob divide there blocks into two parts of the same length, calculate the check bits of two separated blocks, one is the same while the other is different. Then the block whose check bit is different between Alice and Bob will again be divided into two parts of the same length until the only error bit is finally found.
Alice and Bob then independently calculate the check bit of keys and compare the results with each other. During the halving process of error correction in Cascade, a total number of $\log_{2}l_{0}$ bits are transferred for each single block~\cite{Schneier1997Applied}. While the process of privacy amplification will decrease the capacity of keys from $n$ to $m$, so the total number bits transferred for error correction is:
\begin{equation}
l_{EC}=\left[\frac{nN}{ml_{0}}\right] \log_{2}l_{0}.
\end{equation}

So the total number of bits exchanged between Alice and Bob is:
\begin{equation}  \label{con:loadL}
%l=l_{EC}+N=N(1+ \frac{e}{1-e} \frac{n}{r})
l=N+l_{EC}=N+\left[\frac{nN}{ml_{0}}\right] \log_{2}l_{0}.
\end{equation}

%Considering the privacy amplification will decrease the capacity of keys, so the keys generated by one block will not be enough to encrypt one plain text. So the total number of bits communicated when encrypting a plain text of length

Take QBER $e=10\%$ as an example, $l_{0}$ is $[\frac{1}{0.1}]=10$. Assume $\frac{m}{n}=0.2$, it can be calculated from (\ref{con:loadL}) that $l=21.29$ for XOR, $l=170.30 $ for DES, and $l=681.21$ for AES(256).
It can be seen that $l$ increases when using more complex encryption algorithms.

\begin{figure}[h]
	\centering
    \includegraphics[width=1\columnwidth]{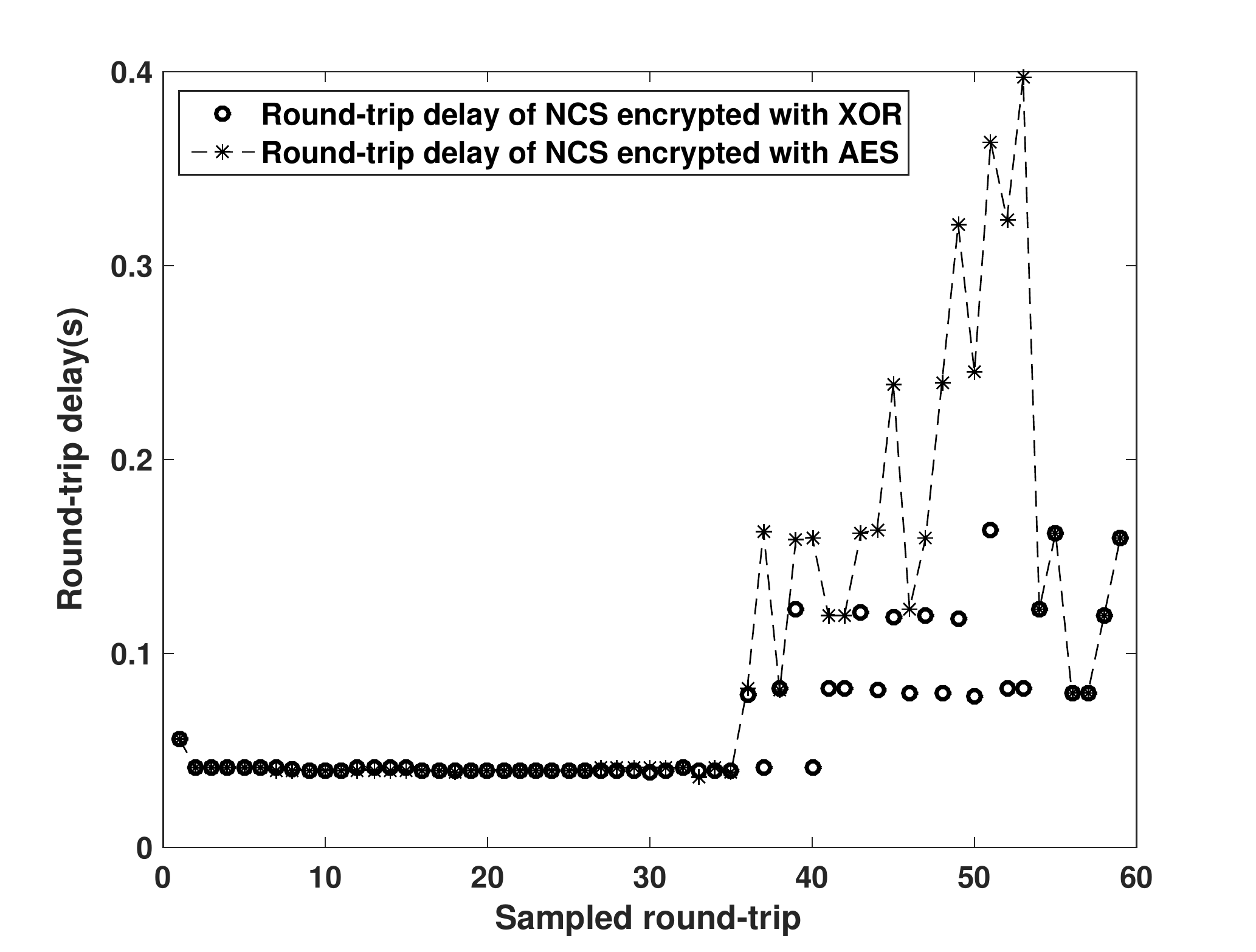}
	\caption{The round-trip delay of NCS encrypted by XOR and AES.}
	\label{fig:SampleDelay}
\end{figure}

To demonstrate how the NCS time delays depend on the complexity of encryption algorithms, we build an experimental networked control servo system and test the XOR and AES algorithms. The round-trip delay is measured by checking the arrival time of two adjacent measurement signals $y(t)$ at Alice's part. As is shown shown in Fig.\ref{fig:SampleDelay}, the experimental round-trip delay increase when the communication load of network starts to be heavier form the 35th control period and the time delay with more complex encryption algorithms (i.e., AES) tend to be longer. The difference is not visible when the communication is not busy at the beginning, but as the experiment goes, complex algorithms collect more data (i.e., cipher text) and consumed more keys. As a result the performance of NCS is degraded.

The round-trip delay of NCS during each control period is highly influenced by the complexity of encryption algorithms, which, according to~\cite{8216956,Misra:2000:FAN:347057.347421}, can be calculated as follows:
\begin{equation}  \label{con:loadtao}
\begin{aligned}
%\tau=T_{0}+\frac{l}{C}=T_{0}+\frac{N}{C}(1+\frac{e}{1-e}\frac{n}{r})+\Delta\tau
\tau &=T_{0}+\frac{2l}{C}+\Delta\tau\\
&=T_{0}+2C^{-1}\left(N+\left[\frac{nN}{ml_{0}}\right] \log_{2}l_{0}\right)+\Delta\tau,\\
\end{aligned}
\end{equation}
where $T_{0}$ is the total time delay without data encryption-decryption and $l/C$ describes the queueing delay. In our networked control system, both the command and realtime position are transmitted through ethernet, then the queueing delay should be doubled. $\Delta\tau$ is the small variable delay caused by hardware, which is much smaller than the delay caused by communication.
%{\color{blue}(\ref{con:loadtao}) means that although there is a stochastic part in the delay of NCS, we can estimate the average value of round-trip delay according to the communication load of different algorithms encrypted with quantum keys.}

The parameters $T_{0}$ and $C$ in (\ref{con:loadtao}) can be identified from the results of the above encrypted NCS experiments using XOR, DES and AES algorithms, which turns out to be: $T_{0}=0.055s$, $C=18000s^{-1}$.

The following theoretical analysis on the tradeoff between the security and the performance will be based on this identified model.

%Based on the formulas above, we can evaluate the induced time delay of NCS according to the designing of encryption algorithms and the conditions of network by choosing values of parameters as: $T_{0}=0.05$, $C=3\times10^{4}$, $e=0.1$, $\frac{n}{r}=5$. The delay for XOR is $\tau=0.05+2\times\frac{21.29}{3\times10^{4}}=0.051s$. For DES, $\tau=0.05+2\times\frac{170.30}{3\times10^{4}}=0.061s$. For 256-bit AES, $\tau=0.05+2\times\frac{681.21}{3\times10^{4}}=0.095s$.%As for the length of keys $N$, it varies according to the encryption

%\begin{table}[htbp]
%\centering
% \caption{Round-trip delay of QKD based NCS encrypted with different algorithms}
% \begin{tabular}{  c  c  c  c  c c }
%  \toprule
%  Encryption algorithms & XOR & DES & RCBC & DES & AES\\
%  \midrule
% $\eta$ & 0.25 & 0.6382 & 0.9072 & 1 & 1 \\
% $l$ & 32 & 32 & 40 & 64 & 128\\
%  \bottomrule
% \end{tabular}
% \label{tab:Comparision}
%\end{table}

\section{Tradeoff Between Security and Performance}\label{Sec:Tradeoff}
The security of QKD-based NCS is determined by the QKD, the algorithm complexity and the one-time pad. According to (\ref{con:securitydefinition}), the compound security can be high even with simple encryption algorithms, so that the length of cipher text can be reduced, which will greatly release the communication burden. As a consequence, the induced time delay caused by classical communication will be reduced as well. In this section, we show how the performance is preserved by simple encryption algorithms without sacrificing too much security.
\subsection{Performance of NCS Encrypted with Different Algorithms}
We evaluate the performance by following the mean square error as:
\begin{equation}
P=\int_{0}^{\infty}e^{2}(t)dt.
\end{equation}

%\begin{figure}[h]
%	\centering
%	\includegraphics[width=0.8\columnwidth]{performance.pdf}
%	\caption{The performance of NCS compared with traditional control system.}
%	\label{fig:compare}
%\end{figure}

%Fig.4 shows about a NCS without encryption and a traditional control system. The measurement from practical product showed that the loop delay of NSC is a random value and we should better make sure that the loop delay is much smaller than sampling period, which will help reform the perform the stability of NCS. By counting the clock of hardware, the loop delay of our networked control system varies from 50ms to 150ms.By calculating the mean square error of two cures with equation(13), the parameter of NCS is 0.063, larger than traditional system, the parameter of which is 0.038.

\begin{figure}[h]
	\centering
    \includegraphics[width=1\columnwidth]{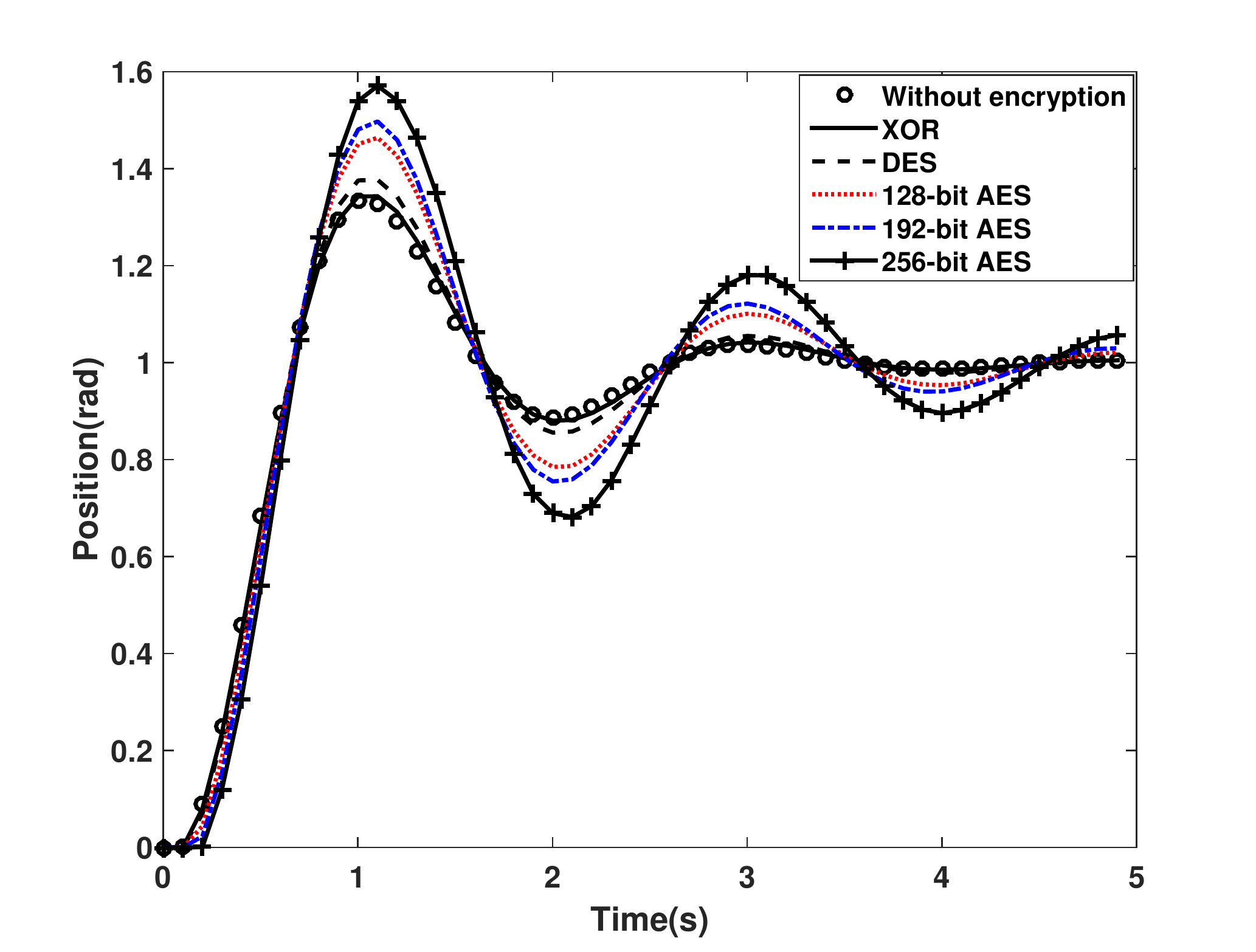}
	\caption{Performance of NCS encrypted with different algorithms.}
	\label{fig:performance}
\end{figure}

%\subsection{The relationship between performance and encryption algorithm}
%In this part we will analyze the delay caused by encryption and decryption.
Secured by QKD, we test the performance under simple and complex encryption algorithms, including XOR, DES and AES. Due to the time delay (\ref{con:loadtao}) induced by the data transmission, the performance with these algorithms are more or less deteriorated, as shown in Fig.\ref{fig:performance}. Both the overshoots and the resetting time increase, in which XOR has the best performance and AES(256) has the worst performance.

As illustrated above, the communication required for QKD increases the time delay and thereby influences the performance. While the time delay caused by encryption and decryption is usually fixed and negligible because the hardware computing power is usually sufficiently high, so the performance is mainly influenced by the induced time delay caused by communication.

\subsection{The Tradeoff between Security and Performance}
%%In this part, we analyze the reformed tradeoff between security and performance with the help of QKD. As showed in equation (18), the value of parameter $\alpha$ will influence the objective function.
%To ensure the performance and security simultaneously, we hope that the normalized parameter of security defined before is relatively high and that of square error is relatively small.
%Based on this, we reform the tradeoff function as follows~\cite{6095624}:
%
% \begin{equation}
%\left\{
%\begin{aligned}
%&f=\alpha S+(1-\alpha)\frac{1}{P}\\
%&0\leq\alpha\leq1\\
%\end{aligned}
%\right.
%\end{equation}

When QKD is used in the NCS, the overall security can be arbitrarily high according to (\ref{con:securitydefinition}) when $\varepsilon$ is sufficiently small. The security parameter is almost independent with the algorithm complexity $S_{A}$. Since the network time delay is shorter when using simple encryption algorithms, there is no need to use complex encryption algorithms that tend to worsen the control performance.

Take $\varepsilon=0.1$ as an example, the security of different encryption algorithms can be enhanced and listed in Table \ref{tab:comparision}.

%\begin{table}[htbp]
%\centering
% %\caption{\label{tab:test}Comarison of perormance of NCS}
% \caption{Comparison of performance and security enhanced by QKD}
% \begin{tabular}{ccccc}
%  \toprule
%  Algorithm  & $S_{A}$  & $S$  & average delay(s) & $P$ \\
%\hline%  \midrule
%  %Null      & 0.050 &  0.0864 & & \\
%  XOR  & 0.0033  & 0.9003             & 0.0432 &  0.3655 \\
% 1-Feistel  & 0.0112  & 0.9011        & 0.061 & 0.3906 \\
% 8-Feistel  & 0.0896  & 0.9089        & 0.061 & 0.3906 \\
% DES  &0.1792   & 0.9179       & 0.08 & 0.3906 \\
% AES(128)   &0.6250   &0.9625         &0.073 & 0.4185 \\
% AES(192)   &0.8121   &0.9812         &0.086 & 0.4443 \\
% AES(256)   &1        &1              &0.095 & 0.4648 \\
%  \bottomrule
% \end{tabular}
%\label{tab:comparision}
%\end{table}

\begin{table}[htbp]
\centering
 %\caption{\label{tab:test}Comarison of perormance of NCS}
 \caption{Comparison of performance and security enhanced by QKD}
 \begin{tabular}{ccccc}
  \toprule
  Algorithm  & $S_{A}$  & $S$  & average delay(s) & $P$ \\
\hline%  \midrule
  %Null      & 0.050 &  0.0864 & & \\
  XOR  &  0.0017  & 0.9002             & 0.0645 &  0.4044 \\
 1-Feistel  & 0.0112  & 0.9011        & 0.0731 & 0.4254 \\
 8-Feistel  & 0.0896  & 0.9089        & 0.0731 & 0.4254 \\
 DES  &0.1792   & 0.9179       &  0.0731 & 0.4254 \\
 AES(128)   &0.6250   &0.9625         &0.0978 & 0.4969 \\
 AES(192)   &0.8121   &0.9812         &0.1118 & 0.5340 \\
 AES(256)   &1        &1              &0.1307 & 0.6300 \\
  \bottomrule
 \end{tabular}
\label{tab:comparision}
\end{table}
%of the diversity of complexity of encryption algorithms will be restrained to be a relative small range, then the adding of encrypting complexity will not influence too much on overall security, so we can decrease $\alpha$ to maximum the objective function. Now we take $\varepsilon=0.1$ as an example:
It can be seen that when encrypted with quantum keys, the security is enhanced no matter which algorithm is used, and the improvement is greater especially for simple algorithms such as XOR and 1-Feistel. The resulting security of different algorithms is much closer to each other. For example, the security of XOR is only about $10\%$ less than that of AES(256), implying that the algorithm complexity is less important. So simple algorithms such as XOR can be used in QKD based network to improve performance without sacrificing the privacy security.

\section{Enhanced Security of NCS with Raw Quantum Key Distribution}\label{Sec:Raw}
As illustrated above, the control performance can be effectively improved by using simple encryption algorithms, while the NCS security level is maintained at a high level. In practice, this requires sufficient supply of secret keys by the QKD system, which may not be realistic because most sifted raw keys are discarded during the post processing (including the error correction and privacy amplification) and the amount of final keys is very limited. The post processing can also decrease the security of the keys if not being well designed.

In this section, we will show that these discarded raw keys can be more efficiently used. This is because the errors of the transmitted data (usually numerical values) caused by the the key errors can be taken as an equivalent noise. Since most closed-loop control processes can tolerate noises at a certain level, the raw keys may not have to be corrected if the equivalent noise is tolerable. In this way, the QKD protocol can be greatly simplified because the succeeding privacy amplification will be also unnecessary, as its function is to remedy the information leakage during error correction. This will bring at least three advantages: (1) the sifted raw keys, which can be fast generated, may be more efficiently used instead of being wasted; (2) the security of keys is the highest without information leakage in error correction; (3) more communication bandwidth can be saved without having to preform the post processing.

To implement the above idea, we need to make sure that the key error induced noise in the transmitted data is suppressible. From the point of view of cryptography, the avalanche effect of the encryption algorithm, namely the sensitivity of the decrypted plain text with the change of the cipher text, should be as weak as possible, otherwise a single one-bit error may cause uncontrollable large deviation. In this regard, XOR is the best choice, under which every one-bit error in the key only affects one bit in the message transmitted from Alice to Bob.

The avalanche effect can be also understood from the control model (\ref{con:controlmodel}). When the key error $\Delta_i$ occurs, the avalanche effect is evaluated by the difference
\begin{equation}\label{}
%w_i=T_{k_i+\Delta_i}^{-1}T_{k_i}u_i-u_i,
w_i=u_i-\bar{u}_i,
\end{equation}
between the original plain text $\bar{u}_i$ and decrypted command $u_i$ after encryption-decryption. It is desired that the encryption matrix $T_{k_i}$ is chosen such that $d_i$ is as insensitive as possible to the key error $\Delta_i$, and here we choose the bit-by-bit XOR encryption. If the random key errors all occur to the lower digits of the plain text, it is reasonable to take the resulting deviation of $u_i$ as a white noise.

Similarly, the error in the key $k'_{i}$ used for the encryption of feedback position of Bob will also lead to an equivalent noise as:
\begin{equation}\label{}
%\nu_i=T_{k'_i+\Delta'_i}^{-1}T_{k'_i}\bar{y}_{i}-\bar{y}_{i}.
\nu_i=y_{i}-\bar{y}_{i}.
\end{equation}

This leads to the following equivalent stochastic control model:
 \begin{equation}
\left\{
\begin{aligned}
&x_{i}=Ax_{i-1}+B(u_{i-1}+\omega_{i-1})\\
&y_{i}=Cx_{i}+\nu_{i},\\
%&z_{i}=T_{k'_i+\Delta'_i}^{-1}T_{k'_i}y_{i}=Cx_{i}+\nu_{i},\\
%&u_{i}=F\cdot T^{-1}_{k_i+\Delta_i}T_{k_i}\cdot z_{i},
%&u_{i}=F(r_{i}-(Cx_{i}+\nu_{i})),
\end{aligned}
\right.
\end{equation}
where $y_i$ is the measurement result that Alice receives through ethernet.

%and $\nu_{i}$ is the equivalent noise induced by key errors $\Delta'$ in the transmission of measurement data encrypted with $k'_i$.

\begin{figure}[!htb]
	\centering
    \includegraphics[width=1\columnwidth]{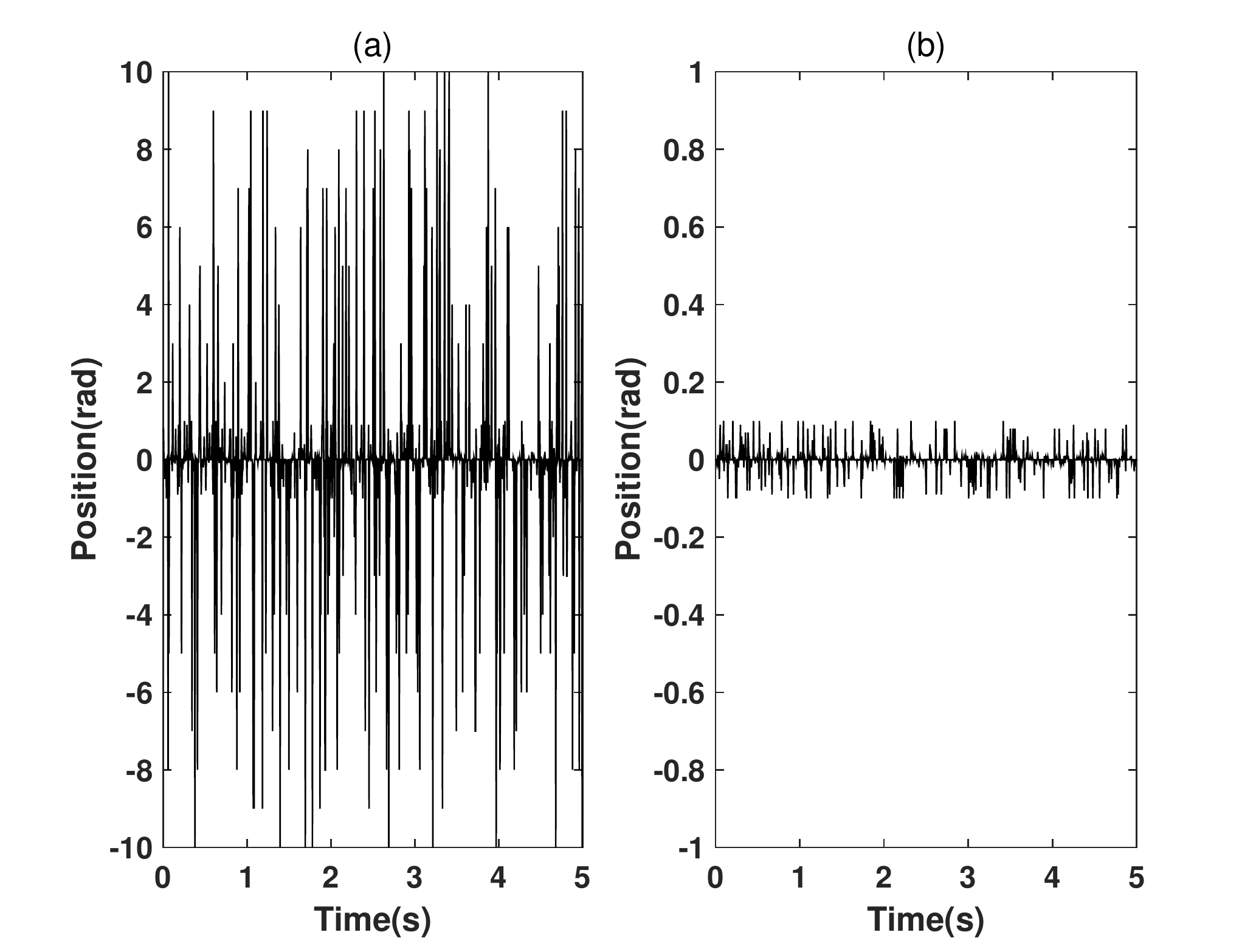}
	\caption{Equivalent noise in the position data of the servo systems that is induced by the errors (with error rate $\approx 10\%$) in the raw keys. (a) the original noise; (b) the noise after removing errors in high digits.}
	\label{fig:ErrorOfKey}
\end{figure}

If the error occurs in the higher digits, the resulting deviation can be filtered out by additional detection and smoothing operations. Concretely, we can check the value of
$d_i=\|y_{i}-y_{i-1}\|$ between adjacent received signals $y_{i}$ and $y_{i-1}$. Once $d_i$ exceeds some prescribed threshold value, say $\delta$, a key error is deemed to occur to the high digits. We can then simply smooth it by making all high digits of $y_i$ identical with those of $y_{i-1}$, which is shown in Fig.~\ref{fig:ErrorOfKey}. Note that this strategy is very effective when the signal to be transmitted is slowly varying. For violently changing signals, one should still use the corrected final keys to encrypt the high digits.

%\begin{equation} \label{con:var}
%\sigma=\sum_{i=L_{0}-L_{H}}^{L-L_{H}}\frac{(10^{-2i})^{2}}{12}\frac{1}{L}.
%\end{equation}
%\begin{equation}
%\phi_{k,k-1}=I+TA+\frac{1}{2}T^{2}A^{2},
%\end{equation}
Now we only need to consider the error in the lower digits that is not detectable with the above method. Take them as white Gaussian noises, whose variance is proportional to the threshold value $\delta$. In the beginning, we use the corrected final keys for encryption-decryption to guarantee that no errors occur to the high digits of the data. Afterwards, we encrypt and decrypt the data all with raw quantum keys, and the noises induced by key errors can be attenuated by the Kalman Filter~\cite{Chen2017Event}. Let $\hat{x}_{i,i-1}$ be the priori estimation of state vector at the end of the $i$-th control period. It can be calculated by
\begin{equation}
\hat{x}_{i,i-1}=A\hat{x}_{i-1}+Bu_{i-1},
\end{equation}
based on the posteriori state estimation $\hat{x}_{i-1}$ and its resulting control command $u_{i-1}=F\cdot(r_{i-1}-C\hat{x}_{i-1})$ at the previous period, following which the posteriori state estimation in the $i$-th control period is made as follows:
\begin{equation}
\begin{aligned}
\hat{x}_{i}=\hat{x}_{i,i-1}+K_{i}(y_{i}-C\hat{x}_{i,i-1}),
\end{aligned}
\end{equation}
where matrices $P_{i,i-1}$, $P_{i}$ and $K_{i}$ are calculated as follows:
\begin{eqnarray}
P_{i,i-1}&=&AP_{i-1}A^{T}+Q,\\
P_{i}&=&(I-K_{i}C)P_{i,i-1},\\
K_{i}&=&P_{i,i-1}C^{T}(CP_{i,i-1}C^{T}+R)^{-1},
\end{eqnarray}
with $Q$ being the covariance matrices of the process noise $\omega_{i}$ and $R$ being the covariance matrices of measurement noise $\nu_{i}$.

\begin{figure}[!htb]
	\centering
    \includegraphics[width=1\columnwidth]{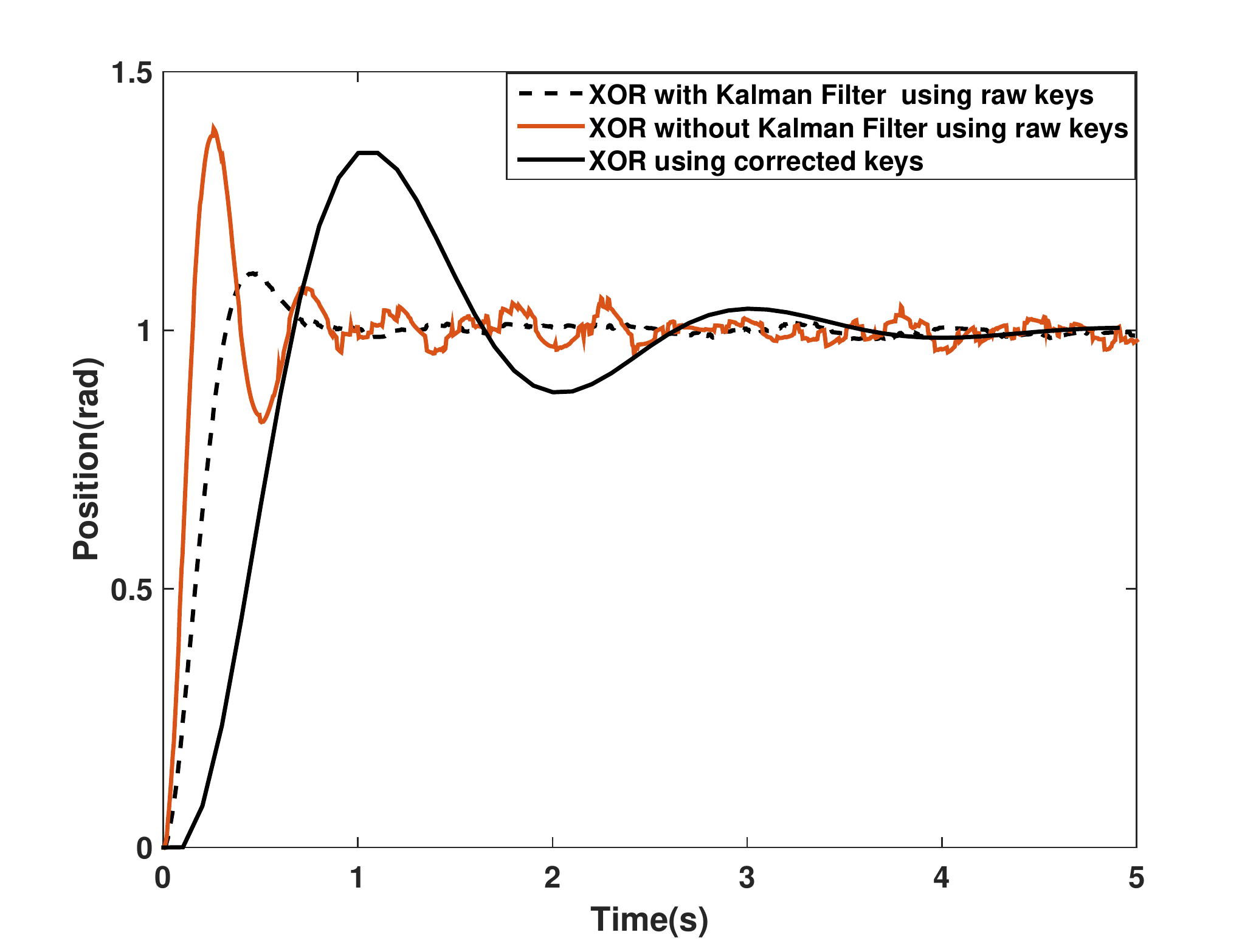}
	\caption{Performance of Raw QKD encrypted NCS with or without Kalman Filter.}
	\label{fig:compare}
\end{figure}

In Fig.~\ref{fig:compare}, we simulate the dynamical response of the NCS with and without the Kalman Filter. Compared with the previous simulation using corrected keys, it can be seen that the use of raw keys introduces additional noises to the output signal, but the transient response is much faster because the induced time delay gets shorter without post processing. The introduction of the Kalman Filter can not only smooth out the fluctuation but also suppress the overshoot, showing that both good control performance and high security can be achieved with raw keys.

\section{Conclusion}\label{Sec:Conclusion}
%To conclude, we proposed a novel scheme for improving the cyber-security of networked control systems by using the one-time pad encryption that is enabled by the emerging technology of quantum key distribution. We analyzed the factors that affect the overall security of NCS, and derived a security measure for the evaluation and design of secure networked control systems. Based on this measure, we studied the tradeoff between security and control performance using various encryption algorithms from the simplest XOR and the most complicated AES, showing that QKD can greatly improve the performance whiling maintaining the security at a high level. To more efficiently use the keys generated by QKD, we further proposed that the Kalman filter can be combined to deal with noises caused by errors in the raw keys.

To conclude, we proposed a novel scheme for improving the cyber-security of networked control systems by using the one-time pad encryption that is enabled by the emerging technology of quantum key distribution. We analyzed the factors that affect the overall security of the NCS, and derived a security measure for the evaluation and design of secure networked control systems. Based on this measure, we studied the tradeoff between the security and the control performance using various encryption algorithms from the simplest XOR and the most complicated AES, showing that QKD can greatly improve the performance whiling maintaining the security at a high level. We further propose a scheme for more efficiently exploiting the raw keys that contain errors, and introduce the Kalman Filter to attenuate the error-induced noises.

To our knowledge, this is the first study that integrated QKD with NCS for security enhancement. The provably secure keys generated by QKD enables the use of one-time pad without having to use complex encryption algorithms. The control framework can be applied to many systems, especially when security is highly demanded. Although the current cost of QKD is high, it can be decreased to be affordable under many critical circumstances.

This work lays a fundamental framework for QKD-based NCS. The extension from point-to-point control to multi-agent control over a complex communication network is more challenging. There are many interesting topics, such as the modeling of the QKD network and evaluation of the overall network security, as well as the following control problems such as consensus. These topics are to be studies in future.

%\section{Appendix}\label{Sec:Conclusion}
\begin{appendices}
\section{Security measure of QKD Based NCS}
%{\color{blue}In addition to its physical state space (e.g, position and velocity of a servo), an encrypted NCS also involves the following space:
To evaluate the security strength of an encrypted NCS, we firstly introduce the following cryptographic terminologies:

{\bfseries (1) Plain text space $\mathcal{M}$}, the set of all plain texts that may be decrypted to be. In our system, it is a combination of practical secure communication data and another subset of possible decrypted data that only occurs under cyber attacks.
%shall represents the value of PI command or the real time position of servo and other symbols, namely $\{u_{i},y_{i},\mathcal{E}\}\subset \mathcal{M}$,
%%\begin{equation}  \label{con:Set}
%%\{u_{i},y_{i},\mathcal{E}\}\subset \mathcal{M},
%%\end{equation}
%where $\mathcal{E}$ is the set of symbols or messy code used for detection of the error of shared keys between Alice and Bob or cyber attacks on the cipher text by checking the decrypted message.

{\bfseries (2) Cipher text space $\mathcal{C}$}, the set of cipher texts after encryption, which will be transmitted through the communication network. %namely $\{T_{K_i}\cdot u_{i},T_{K_i}\cdot y_{i}\}\subset \mathcal{C}$.

{\bfseries (3) Key space $\mathcal{K}$}, the set of secret keys, which are usually binary strings. % namely $K_i\in \mathcal{K}$.

A well defined plain text space will add the difficulty for eavesdropper to crack the message. In our example system, it contains the digitized values of PI command, the real-time position of servo and other symbols, which will enlarge the confusion effect of encryption.

The overall security is mainly determined by the security of keys, the security of key management and the security of encryption algorithms.

\subsection{Security of Keys}
The security of an encryption system can be defined by the mutual information $I(\mathcal{M},\mathcal{C})$ between the plain text space $\mathcal{M}$ and the cipher text space $\mathcal{C}$. Ideally, it should be zero, meaning that the plain text space $\mathcal{M}$ is absolutely independent from the cipher text space $\mathcal{C}$. So Eve is unable to get any meaningful message from the cipher text. In practical system, the mutual information between plain text and cipher text can not be zero but the value can approach to zero after post processing methods.
%This is interlinked with formula (5).

%As for the security of keys, it can be proved that the information that eavesdropper can get after privacy amplification is exponentially small, namely:

The security of QKD is due to the fact that the information about keys that Eve can get is exponentially small. In~\cite{Renner2004Universally}, Renner proposed the composable security of QKD determined by the process of error correction and privacy amplification. So the security discussion on QKD can be divided into following two parts.

\subsubsection{Information leakage of Cascade}

Once the raw keys are distributed by measuring the phase of photons and exchanging the information about bases, there are still some error bits, namely some bits of raw keys are different between Alice and Bob because of the error of measurement. So it is necessary to distill the raw keys and delete the error bits. When Cascade is used in error correction, there is still some information leakage.

\subsubsection{Security of privacy amplification}
Because the fact that Eve can get some formation about sifted keys by tapping the communication of error correction, the process of privacy amplification will repair the leakage by decreasing the key capacity. %This can be ensured by Hash function.

 We assume that Alice and Bob have shared a key string $W$ and the length of $W$ is $n$. Eve has some information about $W$ which can be defined as $V$ and the length of $V$ is $t$. It is reasonable to assume that:
%We use $X$ to represent Alice or Bob, $E$ to represent eavesdropper. It is reasonable to assume that:

\begin{equation}
0<I(\mathcal{K},V)<I(\mathcal{K},W),
\end{equation}
where the information that Eve can get about keys can be calculated by mutual information as $I(\mathcal{K},V)$. Alice and Bob can randomly share a Hash function, namely $g:(0,1)^{n}\rightarrow(0,1)^{m}$. As proved in~\cite{476316}, the mutual information that Eve can get about keys after privacy amplification $I(\mathcal{K},GV)$ is limited as follows:

\begin{equation}
I(\mathcal{K},GV)\leqslant\frac{2^{-n+t+m}}{\ln2}.
\end{equation}

Define $s=n-t-m$, then

\begin{equation}
I(\mathcal{K},GV)\leq\frac{2^{-s}}{\ln2},
\end{equation}
where $s$ represents the reduction ratio of the capacity of keys during privacy amplification.
As a result, the information that Eve can get from keys after privacy amplification is exponentially small.

\subsection{Security of Key Management}
One-time pad is the only provable absolutely secure encryption method even under unlimited computation power. The security of one-time pad has two aspects. On the one hand, give a copy of cipher text, the eavesdropper gains no more information about the plain text even after trying every possible key, namely,

\begin{equation}
P[M=m_{0}|C=c]=P[M=m_{0}],
\end{equation}
which can be called information-theoretic security~\cite{Schneier1997Applied}.

On the other hand, when one-time pad cannot be ensured due to the limited supply of keys, some keys may be reused, this circumstance can be described as:

% \begin{equation}
%\left\{
%\begin{aligned}
%m_{i}&\bigoplus k_{i} =c_{i}\\
%m_{i+(r-1)}&\bigoplus k_{i}=c_{i+(r-1)}\\
% \vdots &\hspace{1cm} \vdots\\
%m_{i+r}&\bigoplus k_{i}=c_{i+r}\\
%\end{aligned}
%\right.
%\end{equation}

\begin{equation}
m_{i+(r-1)}\oplus k_{i}=c_{i+(r-1)},
\end{equation}
where $r$ is the times that a single key $k_{i}$ is reused. The information that Eve can get from the cipher text on the plain text can then be valued by the conditional probability. Namely on knowing the cipher text is $C=c_{i+1}$, the probability that plain text is $M=m_{i+1}$ can be calculated according to the Bayes Rule as:

\begin{equation}  \label{con:OTPr}
\begin{aligned}%{split}
&P[M=m_{i+1}|C=c_{i+1}]= \frac{P[M=m_{i+1}\bigcap C=c_{i+1}]}{P[C=c_{i+1}]}\\
&= \frac{P[C=c_{i+1}|M=m_{i+1}]P[M=m_{i+1}]}{\sum_{m\in M}P[C=c_{i+1}|M=m]P[M=m]}.\\
%&= \frac{r 2^{-l}P[M=m_{i+1}]}{2^{-l}}
\end{aligned}%{split}
\end{equation}

Considering the difficulty for eavesdropper to crack the meaning of message, the whole key space should be tried until a meaningful message is calculated. Define the length of plain text and key are both $l$, then there are $2^{l}$ probable messages uniformly distributed in the space. When a single key is reused for $r$ times, the information available for eavesdropper will be added by solving (\ref{con:OTPr}). When connected with Bayes formula, the conditional probability $P[C=c_{i+1}|M=m_{i+1}]$ will be $r$ times larger because of the reusing of key $k_{i}$. Then the information that Eve can get from cipher text on plain text can be:

\begin{equation}
\begin{aligned}%{split}
P[M=m_{i+1}|C=c_{i+1}] &= \frac{r 2^{-l}P[M=m_{i+1}]}{2^{-l}}\\
&=rP[M=m_{i+1}] \\
&=r2^{-l}.
\end{aligned}%{split}
\end{equation}
Additionally we assume that the times a single key is reused is limited as $r<2^{l}$, other then the key can be regarded as unchangeable.

The derivation shows that when the keys are reused, the information that Eve can get from cipher text is increased, which degrades the security of practical systems.
%P[M=m_{i+1}|C=c_{i+1}]&=\frac{P[M=m_{i+1}\bigcap C=c_{i+1}]}{P[C=c_{i+1}]}\\
 %   &=\frac{P[C=c_{i+1}|M=m_{i+1}]P[M=m_{i+1}]}{1}
%C=c_{i+1}]}{P[C=c_{i+1}]}=\frac{P[C=c_{i+1}|M=m_{i+1}]P[M=m_{i+1}]}{\sum_{m\inM}P[C=c_{i+1}|M=m]P[M=m]}=\frac{r2^{-l}P[M=m_{i+1}]}{2^{-l}}=rP[M=m_{i+1}]

The security of one-time pad is also influenced by the updated keys. While in our system, the randomness is well ensured by the physically randomly generated single photons.

\subsection{Security of Encryption Algorithms}
To define the security of encryption algorithms, the parameters should be reasonable rather than measured from the cipher text, and they should be connected with the complexity of algorithms and the difficulty for brute force. Avalanche effect and linear cryptanalysis, which are measured according to the results of cipher text, can be easily affected by the choosing of plain text and keys. So we use the following parameters to define the security of encryption algorithms:

\subsubsection{Length of keys $N$} The length of keys will influence the number of keys that should be used for brute force. For example, if the length of keys is $N$, then there will be $2^{N}$ keys need to be tried. While this is the maximum time of testing for brute force attack. The security of encryption should be defined according to the most efficient method of cracking and considering usually concurrent computation is used in hardware, we use $\log_{2}N$ to define the security contributed by the length of keys and this definition is introduced more in detail in~\cite{Haleem2007Opportunistic}.

\subsubsection{Number of rounds $R$} Most encryption algorithms have its own round function, for example DES is 16 rounds of Feistel, and the adding of rounds will increase the complexity of encryption.

\subsubsection{Complexity of round operation} Round operation is the encrypting unit of a complex encryption algorithm, and it will encrypt the plain text by XOR with the keys, substituting bytes, shifting rows, mixing columns and so on. All the operations are reversible and will change the states of plain text.

%All the operations can be conveyed by S-box, which can represent the rules of encryption.*****************

The security of round operations is a combination of the security of subkeys and the security of calculation operations. According to key expanding algorithm, a subkey is decided by that of upper round and updating method, and the latter is public. So the security of subkey is equivalent to the security of original keys, namely the security of quantum key distribution, which has been defined before. Based on this, the security of round function is equivalent to the security of calculation operations.

According to Table \ref{tab:AlgorithmComparision}, we can define the complexity of round function $\eta$ by adding the security of ByteSub $S_{B}$, the security of ShiftRow $S_{S}$, the security of MixColumn $S_{M}$ and the security of RoundKeyAddition $S_{R}$ as follows:

\begin{equation}
\eta=S_{B}+S_{S}+S_{M}+S_{R}.
\end{equation}

Now we can come up with the following conclusions:

(a) ByteSub can enhance the security by substituting the plain text according to S-box, a designed method mapping the bytes of plain text to the elements of S-box. The security of ByteSub is influenced by the size of S-box.

(b) ShiftRow, MixColumn and RoundKeyAddition are three methods of changing the value of bits and they are equivalent when functioning on the plain text. While the complexity can be different because of the designed rule and that will influence the security.

To quantitatively compare the difference of varied encryption algorithms, we assume the complexity of SPN is 1 composed by four parts of the same weight and define the complexity of other round functions by their relative complexity of operations, namely the time of operations compared with SPN according to TABLE \ref{tab:AlgorithmComparision}, namely,
\begin{equation}
\eta(*)=\sum\frac{C^{\ast}}{C_{SPN}}S_{SPN},
\end{equation}
where $C_{SPN}$ means the complexity of AES (SPN), $C^{\ast}$ means the complexity of other algorithms.

\begin{table}[htbp]
\centering
 \caption{Comparison of the times of operations and security for round functions of different encryption algorithms}
 \begin{tabular}{  c | c | c | c }
  \toprule
  Algorithms &  XOR & Feistel (DES) & SPN (AES) \\
\hline %  \midrule
  ByteSub & 0 &  $4\times16$  & $16\times16$ \\
  ShiftRow & 0 &1 & 3 \\
  MixColumn & 0 &0 & 1 \\
  RoundKeyAddition & 32 & 32& 128 \\
\hline%  \midrule
  $S_B$ & 0 & 0.0625 & 0.25   \\
  $S_S$ & 0 & 0.0833 & 0.25   \\
  $S_M$ & 0 & 0 & 0.25   \\
  $S_R$ & 0.0625 & 0.0625 & 0.25   \\
  $\eta$ & 0.0625  & 0.2083 &  1  \\
  \bottomrule
 \end{tabular}
 \label{tab:AlgorithmComparision}
\end{table}

%For SPN:
%
%\begin{equation}
%\left\{
%\begin{array}{l}
%S_{B}=S_{S}=S_{M}=S_{R}=0.25\\
%\eta=S_{B}+S_{S}+S_{M}+S_{R}=1\\
%\end{array}
%\right.
%\end{equation}
%
%For Feistel:
%
%\begin{equation}
%\left\{
%\begin{array}{l}
%S_{B}=\frac{4\times16}{16\times16}\times0.25=0.0625\\
%S_{S}=\frac{1}{3}\times0.25=0.0833\\
%S_{M}=0\\
%S_{R}=\frac{32}{128}\times0.25=0.0625 \\
%\eta=S_{B}+S_{S}+S_{M}+S_{R}=0.2083\\
%\end{array}
%\right.
%\end{equation}
%
%For XOR:
%
%\begin{equation}
%\left\{
%\begin{array}{l}
%S_{B}=0\\
%S_{S}=0\\
%S_{M}=0\\
%S_{R}=\frac{64}{128}\times0.25=0.125 \\
%\eta=S_{B}+S_{S}+S_{M}+S_{R}=0.125\\
%\end{array}
%\right.
%\end{equation}

%So the security of SPN, Feistel and XOR can be valued as follows:
%
%\begin{table}[htbp]
%\centering
% \caption{Comparing the security of different encryption algorithms}
% \begin{tabular}{  c  c  c  c  c }
%  \toprule
%  Algorithms & $\eta$ \\
%  \midrule
% XOR & 0.0125  \\
% Feistel & 0.2083\\
% SPN & 1\\
%  \bottomrule
% \end{tabular}
%\end{table}

%We use parameter $\eta$ to represent the percentage of bits may get changed in a single round function. For XOR, $\eta=1$ because the plain text can be encrypted with the key of the same length. For DES or Feistel, $\eta=0.5$ because it can only encrypt half of the plain text in each round. For AES, $\eta=1$ because it can encrypted the whole text by combining shifting rows and mixing columns.

So we can define the security of encryption algorithms as:
\begin{equation}
S_{A}=\eta R\log_{2}N.
\end{equation}

Based on the analysis above, we give the security definition of QKD based NCS as (\ref{con:SA})(\ref{con:securitydefinition}).

\end{appendices}

\bibliographystyle{IEEEtran}
\bibliography{refs}

% Generated by IEEEtran.bst, version: 1.14 (2015/08/26)
\begin{thebibliography}{10}
\providecommand{\url}[1]{#1}
\csname url@samestyle\endcsname
\providecommand{\newblock}{\relax}
\providecommand{\bibinfo}[2]{#2}
\providecommand{\BIBentrySTDinterwordspacing}{\spaceskip=0pt\relax}
\providecommand{\BIBentryALTinterwordstretchfactor}{4}
\providecommand{\BIBentryALTinterwordspacing}{\spaceskip=\fontdimen2\font plus
\BIBentryALTinterwordstretchfactor\fontdimen3\font minus
  \fontdimen4\font\relax}
\providecommand{\BIBforeignlanguage}[2]{{%
\expandafter\ifx\csname l@#1\endcsname\relax
\typeout{** WARNING: IEEEtran.bst: No hyphenation pattern has been}%
\typeout{** loaded for the language `#1'. Using the pattern for}%
\typeout{** the default language instead.}%
\else
\language=\csname l@#1\endcsname
\fi
#2}}
\providecommand{\BIBdecl}{\relax}
\BIBdecl

\bibitem{4519604}
E.~A. Lee, ``Cyber physical systems: Design challenges,'' in \emph{2008 11th
  IEEE International Symposium on Object and Component-Oriented Real-Time
  Distributed Computing (ISORC)}, May 2008, pp. 363--369.

\bibitem{Kogiso2015}
K.~Kogiso and T.~Fujita, ``Cyber-security enhancement of networked control
  systems using homomorphic encryption,'' in \emph{2015 54th IEEE Conference on
  Decision and Control (CDC)}, Dec 2015, pp. 6836--6843.

\bibitem{510638}
S.~Goto, M.~Nakamura, and N.~Kyura, ``Accurate contour control of mechatronic
  servo systems using gaussian networks,'' \emph{IEEE Transactions on
  Industrial Electronics}, vol.~43, no.~4, pp. 469--476, Aug 1996.

\bibitem{Ding2017A}
D.~Ding, Q.~L. Han, Y.~Xiang, X.~Ge, and X.~M. Zhang, ``A survey on security
  control and attack detection for industrial cyber-physical systems,''
  \emph{Neurocomputing}, 2017.

\bibitem{Kundur2011Towards}
D.~Kundur, X.~Feng, S.~Mashayekh, S.~Liu, and T.~Zourntos, ``Towards modelling
  the impact of cyber attacks on a smart grid,'' \emph{International Journal of
  Security \& Networks}, vol.~6, no.~1, pp. 2--13, 2011.

\bibitem{Teixeira2010Networked}
A.~Teixeira, H.~Sandberg, and K.~H. Johansson, ``Networked control systems
  under cyber attacks with applications to power networks,'' in \emph{American
  Control Conference}, 2010, pp. 3690--3696.

\bibitem{Kailkhura2015Distributed}
B.~Kailkhura, Y.~S. Han, S.~Brahma, and P.~K. Varshney, ``Distributed bayesian
  detection in the presence of byzantine data,'' \emph{IEEE Transactions on
  Signal Processing}, vol.~63, no.~19, pp. 5250--5263, 2015.

\bibitem{Deng2017Defending}
R.~Deng, G.~Xiao, and R.~Lu, ``Defending against false data injection attacks
  on power system state estimation,'' \emph{IEEE Transactions on Industrial
  Informatics}, vol.~13, no.~1, pp. 198--207, 2017.

\bibitem{Shames2015A}
I.~Shames, H.~Sandberg, and K.~H. Johansson, ``A secure control framework for
  resource-limited adversaries,'' \emph{Automatica}, vol.~51, no.~C, pp.
  135--148, 2015.

\bibitem{WANG20162451}
D.~Wang, Z.~Wang, B.~Shen, F.~E. Alsaadi, and T.~Hayat, ``Recent advances on
  filtering and control for cyber-physical systems under security and resource
  constraints,'' \emph{Journal of the Franklin Institute}, vol. 353, no.~11,
  pp. 2451 -- 2466, 2016.

\bibitem{Mathur2015Solving}
R.~Mathur, S.~Agarwal, and V.~Sharma, ``Solving security issues in mobile
  computing using cryptography techniques — a survey,'' in
  \emph{International Conference on Computing, Communication \& Automation},
  2015, pp. 492--497.

\bibitem{Mohd2015A}
B.~J. Mohd, T.~Hayajneh, and A.~V. Vasilakos, ``A survey on lightweight block
  ciphers for low-resource devices: Comparative study and open issues,''
  \emph{Journal of Network \& Computer Applications}, vol.~58, no.~C, pp.
  73--93, 2015.

\bibitem{Deng2017A}
Y.~Deng and X.~H. Cheng, ``A light-weight data encryption mechanism in wireless
  sensor networks,'' \emph{Computer Engineering \& Science}, 2017.

\bibitem{Bernstein2017Post}
D.~J. Bernstein and T.~Lange, ``Post-quantum cryptography,'' \emph{Nature},
  vol. 549, no. 7671, p. 188, 2017.

\bibitem{MR0032133}
C.~E. Shannon, ``Communication theory of secrecy systems,'' \emph{Bell System
  Tech. J.}, vol.~28, pp. 656--715, 1949.

\bibitem{Schneier1997Applied}
B.~Schneier, ``Applied cryptography, second edition : protocols, algorithms,and
  source code in c,'' \emph{Government Information Quarterly}, vol.~13, no.~3,
  p. 336, 1997.

\bibitem{Bennett1984}
C.~H. Bennett and G.~Brassard, ``Quantum cryptography: Public key distribution
  and coin tossing,'' \emph{In Proceedings of IEEE International Conference on
  Computers, Systems and Signal Processing, volume 175, page 8.New York}, 1984.

\bibitem{476316}
C.~H. Bennett, G.~Brassard, C.~Crepeau, and U.~M. Maurer, ``Generalized privacy
  amplification,'' \emph{IEEE Transactions on Information Theory}, vol.~41,
  no.~6, pp. 1915--1923, Nov 1995.

\bibitem{Shor2000Simple}
P.~W. Shor and J.~Preskill, ``Simple proof of security of the bb84 quantum key
  distribution protocol,'' \emph{Phys.rev.lett}, vol.~85, no.~2, p. 441, 2000.

\bibitem{Scarani2008Quantum}
V.~Scarani and R.~Renner, ``Quantum cryptography with finite resources:
  unconditional security bound for discrete-variable protocols with one-way
  postprocessing.'' \emph{Physical Review Letters}, vol. 100, no.~20, p.
  200501, 2008.

\bibitem{Benor2005The}
M.~Benor, M.~Horodecki, D.~W. Leung, D.~Mayers, and J.~Oppenheim, ``The
  universal composable security of quantum key distribution,'' in
  \emph{International Conference on Theory of Cryptography}, 2005, pp.
  386--406.

\bibitem{119991887739}
\BIBentryALTinterwordspacing
S.~J. Lomonaco, ``A quick glance at quantum cryptography,'' \emph{Cryptologia},
  vol.~23, no.~1, pp. 1--41, 1999. [Online]. Available:
  \url{https://doi.org/10.1080/0161-119991887739}
\BIBentrySTDinterwordspacing

\bibitem{Kollmitzer2010Applied}
C.~Kollmitzer and M.~Pivk, ``Applied quantum cryptography,'' \emph{Lecture
  Notes in Physics}, vol. 797, 2010.

\bibitem{Tang2010Research}
S.~Tang and X.~Ma, ``Research of typical block cipher algorithms,'' in
  \emph{International Conference on Computer, Mechatronics, Control and
  Electronic Engineering}, 2010, pp. 319--321.

\bibitem{10.1007/978-3-540-39887-5_19}
S.~Park, S.~H. Sung, S.~Lee, and J.~Lim, ``Improving the upper bound on the
  maximum differential and the maximum linear hull probability for spn
  structures and aes,'' in \emph{Fast Software Encryption}, T.~Johansson,
  Ed.\hskip 1em plus 0.5em minus 0.4em\relax Berlin, Heidelberg: Springer
  Berlin Heidelberg, 2003, pp. 247--260.

\bibitem{8216956}
Y.~Lin, J.~Wang, Q.~L. Han, and D.~Jarvis, ``Distributed control of networked
  large-scale systems based on a scheduling middleware,'' in \emph{IECON 2017 -
  43rd Annual Conference of the IEEE Industrial Electronics Society}, Oct 2017,
  pp. 5523--5528.

\bibitem{Misra:2000:FAN:347057.347421}
V.~Misra, W.-B. Gong, and D.~Towsley, ``Fluid-based analysis of a network of
  aqm routers supporting tcp flows with an application to red,'' \emph{SIGCOMM
  Comput. Commun. Rev.}, vol.~30, no.~4, pp. 151--160, Aug. 2000.

\bibitem{Chen2017Event}
W.~Chen, D.~Shi, J.~Wang, and L.~Shi, ``Event-triggered state estimation:
  Experimental performance assessment and comparative study,'' \emph{IEEE
  Transactions on Control Systems Technology}, vol.~PP, no.~99, pp. 1--8, 2017.

\bibitem{Renner2004Universally}
R.~Renner and R.~König, ``Universally composable privacy amplification against
  quantum adversaries,'' \emph{Lecture Notes in Computer Science}, vol. 3378,
  pp. 407--425, 2004.

\bibitem{Haleem2007Opportunistic}
M.~A. Haleem, C.~N. Mathur, R.~Chandramouli, and K.~P. Subbalakshmi,
  ``Opportunistic encryption: A trade-off between security and throughput in
  wireless networks,'' \emph{IEEE Transactions on Dependable \& Secure
  Computing}, vol.~4, no.~4, pp. 313--324, 2007.

\end{thebibliography}

\end{document}